\documentclass[12pt]{iopart}
\usepackage{amssymb,latexsym,mathrsfs}
\usepackage{amsfonts}
\usepackage{graphicx}
\newcommand{\bx}[1]{  \textbf{[}#1\textbf{]}}

\newcommand{\beq}{\begin{equation}}
\newcommand{\eeq}{\end{equation}}
\newcommand{\bearr}{\begin{array}}
\newcommand{\enarr}{\end{array}}

\newcommand{\ket}[1]{|#1\rangle}
\newcommand{\bra}[1]{\langle#1|}
\newcommand{\ra}{\rangle}
\newcommand{\la}{\langle}

\def\bea{\begin{eqnarray}}
\def\eea{\end{eqnarray}}
\def\ba{\begin{array}}
\def\ea{\end{array}}
\def\n{\nonumber} 
\def\c{\mathscr}

\begin{document}
\title{Spatial Correlations
  in Exclusion Models corresponding  to the Zero Range Process}
\author{ Urna Basu and P. K. Mohanty}

\address{Theoretical Condensed Matter Physics Division, Saha Institute of Nuclear Physics,
1/AF Bidhan Nagar, Kolkata, 700064 India.}
\ead{urna.basu@saha.ac.in}

\begin{abstract}
We show that the  steady state weights of all one dimensional 
exclusion  models which are mapped to the Zero Range Process (ZRP) can  be written 
in a matrix product form, where the  required matrices 
depend only on the steady state weights of ZRP. One infinite dimensional representation 
of these matrices which works for generic systems has also been provided. This is in contrast 
to the usual matrix product ansatz which does not always guarantee a solution for the 
dynamics dependent algebra that the matrices need to satisfy.
The formulation helps us study the spatial correlations of these exclusion processes 
which are unreasonably difficult to obtain directly from their ZRP correspondence. To illustrate 
this method we  reproduce certain known results, and 
then investigate unexplored correlations in some other model systems. 
\end{abstract}
\maketitle

One dimensional driven diffusive systems with stochastic dynamics, motivated on 
physical grounds, are studied to understand  macroscopic properties of 
nonequilibrium steady states \cite{DDS}. These systems set a ground for the study 
of exotic correlations present in  generic non-equilibrium systems \cite{book}. 
Due to  lack of generic methods
analytical studies are limited to only those few systems  
\cite{ZRP, TASEP, AHR, KLS, Jain, RASEP} for which certain specific techniques 
can be applied. The zero range process (ZRP) is one such  exactly solvable 
model which displays rich non-equilibrium behaviour,
despite of having a simple factorized steady state. Several 
one dimensional systems  which are used to model
dynamics of avalanches, granular systems, interface growth, polymer dynamics, 
various transport processes, and glasses \cite{ZRPrev} can be mapped to ZRP.

In  ZRP, particles from one box hop to a neighbouring box 
with a rate that depends on the number of particles at the departure box. 
The boxes  are either vacant or can accommodate  arbitrary number of 
indistinguishable particles ignoring  hard-core interaction. Several lattice 
 models \cite{AHR, RASEP, Levi04, EKLS} with hard-core interaction ($i.e.$ exclusion processes)
can be mapped to ZRP when  vacant and occupied sites are identified 
as boxes and particles respectively.  These  exclusion processes  have  the same 
steady state  weights as that of the corresponding ZRPs. However it is practically 
impossible to calculate the 
spatial correlations on the lattice  as they are extremely complicated functions of ZRP variables.

 In this Letter, we introduce an exclusion process  on a one dimensional 
ring  with the dynamics that is equivalent to a generic ZRP. To calculate
the spatial correlations  we provide a formulation where  steady state of 
the lattice system can be expressed in a matrix product form;   
the particles and holes  are represented by two  
non-commuting  matrices. A  
specific representation of these matrices, which depends only on 
the steady state weight of the corresponding ZRP, is provided. 
The  formulation can be generically used to study spatial correlations
of exclusion processes having a ZRP correspondence, which is 
illustrated with  a few examples.

The exclusion process is defined on a   one dimensional periodic lattice 
where the sites labeled by $i= 1\dots L$ can accommodate at best one  hard-core particle. 
Any  arbitrary configuration of the system  
is  denoted by $\{s_i\}$
where $s_i=0,1$ corresponds to vacant and occupied sites respectively. 
Particles  hop to a neighbouring vacant site, say to its right, according to 
the following dynamics,  
\beq
0\bx{n} 0 \to 0\bx{ n-1} 01  {\rm  ~with  ~ rate~}  u(n),
\label{eq:rate} 
\eeq
where a block of size $n$, $i.e.$ an uninterrupted  sequence of $n$ $1$s, is denoted by $\bx n$. 
Here, the total number of particles $N$  and thus particle density $\rho = {N \over L}$  remains 
conserved. This dynamics does not necessarily indicate  the presence of  a long range interaction 
in the system. In fact several one dimensional driven systems \cite{AHR,EKLS}, where particles 
having  extra degrees of freedom interact through a short range dynamics, can be  described 
effectively   by Eq. (\ref{eq:rate}).

This exclusion process is equivalent 
to a ZRP with $N$ particles distributed in $M=L-N$ boxes with hop rate $u(n)$, 
provided  $0$s and $1$s are referred to as boxes  and particles respectively. 
 
The steady state  weight  $P_{ZRP} (n_1,n_2 \dots n_M)$ of  a configuration $\{n_i\}$  in ZRP  
has a product measure  for  any generic rate $u(n)$ 
\beq
P_{ZRP} (n_1,n_2 \dots n_M) =  \prod_{k=1}^Mf(n_k)  
\label{eq:sszrp}
\eeq 
with $f(n) =   \prod_{l=1}^n u(l)^{-1}$ being the weight of a box having 
$n$ particles \cite{ZRP}. Equation (\ref{eq:sszrp}) can be used to calculate the 
weight of the corresponding configuration of the exclusion process as follows.
Since the system is  both periodic and translationally 
invariant, the weight $P(\{s_i\})$ of a configuration $\{s_i\}$   
can always be   written as  $P(0\bx{n_1} 0\bx{n_2} \dots 0\bx{n_M})$ where  $\bx {n_k}$ is 
the block after $k^{th}$ $0$ on the lattice. From this correspondence 
it is evident that 
\bea
P(\{s_i\})&\equiv& P(0[n_1]\dots 0[n_M]) \label{eq:eqv1} 
=P_{ZRP} (n_1,\dots n_M)= \prod_{k=1}^M f(n_k).
\label{eq:eqv}
\eea

To obtain the spatial correlation functions of the exclusion process 
from the  mapping to ZRP, one needs to express the site variables $s_i$ in 
terms of the occupation numbers $\{n_k\}$.
\bea
s_i =  1- \sum_{k=1}^M \delta_{i,g_{_k}}~~;\;~~ g_{_k} =  \sum_{l=1}^k  (1+n_l). 
\eea
Clearly,  even the two point correlation functions $\langle s_{i} s_{i+j}\rangle$ 
for arbitrary $j$  are hardly possible to calculate. 
Alternatively  let us  rewrite the steady state weight of the exclusion 
process on the lattice as 
\beq 
P(\{s_i\}) = Tr \left[   X_{s_1} X_{s_2} \dots X_{s_L}  \right],
\label{eq:Tr}
\eeq 
where non-commuting matrices $ X_{s_i}$ keep track of the position indices.
This form is similar to the steady state weights assumed in Matrix 
Product Ansatz (MPA) \cite{MPA}. However, $X_{s_i}$ here satisfy a 
unique relation [Eq. (\ref{eq:main})  below] which depends only on the 
single box weight $f(n)$ of the corresponding ZRP whereas MPA-matrices need to 
fulfill different algebraic relations depending on the concerned dynamics. 

For simplicity, we use   $X_0 \equiv E$ and $X_1\equiv D$.   
From Eqs. (\ref{eq:eqv1}) and (\ref{eq:Tr})  we have 
\beq
P(\{s_i\})= Tr \left[ED^{n_1} ED^{n_2}\dots ED^{n_M}  \right],
\label{eq:ED}
\eeq
which can be simplified further by taking 
$E = \ket \alpha   \bra \beta$, where $ \ket \alpha $ and   $\bra \beta$  
are yet to be determined.  This choice of $E$, along with Eqs. (\ref{eq:eqv})
and (\ref{eq:ED}), demands that $D$ must satisfy the following relation 
\beq
 \bra \beta  D^n \ket \alpha  = f(n).
\label{eq:main} 
\eeq
This equation does not specify $\ket \alpha  $, $\bra \beta$  and $D$ uniquely. 
For this formulation to work  it is sufficient to have  one  
suitable choice of  $\ket \alpha  $, $\bra \beta$  and $D$  which
satisfy the above equation. A finite dimensional representation of these matrices
is guaranteed only for certain special $f(n)$, $e.g.$ when  $f(n)$ is periodic 
 in $n$, or when $f(n) = c^n$.  For generic $f(n)$, however,
one can use the following infinite-dimensional representation   
\bea
 D  = \sum_{i=1}^\infty \ket i \bra {i+1} ~; ~~~~
 \bra \beta = \bra 1,   ~ ~~ {\rm  and}~~~  
\ket \alpha =    \sum _{i=1}^\infty f(i-1) \ket i , 
\label{eq:rep}
\eea 
where $ \{  \ket i \} $   is the  standard basis in infinite dimension, $i.e.$, $ \ket i_k  = \delta_{i,k}$. 
Both $\bra \beta$  and $D$  are independent of the systems concerned;  $\bra \beta$  is a constant vector and  $D$ is a shift operator which has  the following properties
\beq 
Tr[D^m]=0 ;~~{\rm and}~  D^m  \sum_{i=1}^\infty  a_i \ket i =  \sum_{i=1}^\infty a_{i+m} \ket i,  
\label{eq:Dm}
\eeq
where $\{a_i\} $ are arbitrary coefficients. It is only $\ket \alpha$  which depends on the specific 
process, and can be constructed directly from the set $\{f(n)\}$.
Thus, the representation (\ref{eq:rep}) works for \textit{any} exclusion  process  which  can be 
mapped to ZRP; even when $f(n)$ does not have any functional dependence on $n$.

The first task is to calculate the partition function. This is done in the grand canonical ensemble 
(GCE) by  associating the fugacity $z$ with particles ($D$) resulting in $Z_L(z) =Tr[C^L]$ where 
$C=zD +E$. The weight of the configuration having no vacant site  is $Tr[D^L]=0$.
Thus,  $Z_L(z)$ is the sum of the weights of all other configurations with at least one vacant site.
\bea
 Z_L(z) =   \sum_{n=1}^L \Tr\left[  (zD)^ {n-1} E C^{L-n}\right]=\sum_{n=1}^L\bra \beta C^{L-n} (zD)^ {n-1} \ket \alpha.
\label{eq:ZL}
\eea
 The  partition function  can be calculated explicitly 
for some simple forms of $f(n)$. 
For arbitrary  $f(n)$, however,  it is better to use the following generating function  
\bea
\c Z(z,\gamma)=  \sum_{L=1}^\infty \gamma^L Z_L(z)
= \bra \beta\frac{\gamma}{{\cal I}-\gamma C}  \frac{1}{{\cal I}-\gamma z D}  \ket \alpha,
\label{eq:Zzg}
\eea
where  $Z_L$ is the  coefficient of $\gamma^L$  in the series 
expansion of   $\c Z(z,\gamma)$.  Instead,  $\c Z(z,\gamma)$ can be interpreted 
as  the partition function of the system  in variable length ensemble (VLE).
Together $z$ and $\gamma$, here,  determine the macroscopic variables 
\bea
\la N\ra = \frac{z}{\c Z} \frac{\partial \c Z} {\partial z} ~~~~ {\rm and}~~~~
\la L\ra = \frac{\gamma}{\c Z} \frac{\partial \c Z}{ \partial \gamma }.
\label{eq:LN}
\eea
In the thermodynamic limit  $\la L\ra \to \infty$, where  VLE  is expected to be 
equivalent to GCE, one of the variables (say $\gamma$)  becomes a function of 
the other one ($z$). Details of this equivalence will be discussed elsewhere.

Any $n$-point correlation function  can be  expressed  easily in 
terms of holes denoted by $\bar s_i=1-s_i$. 
For  example $\la \bar s_i\ra = \sum_L Tr[ (\gamma E) (\gamma C)^{L-1}]$, can be  written as 
\bea
\la \bar s_i\ra &=& G^{(1)}(z,\gamma)    =   {\gamma \over \c Z}\bra \beta\frac{1}{{\cal I}-\gamma C}\ket \alpha. \label{eq:G1}
\eea
The density $\rho(z,\gamma)= 1- G^{(1)}(z,\gamma)$, obtained from above, is same as that 
calculated using Eq. (\ref{eq:LN}) in the thermodynamic limit.
Since spatial indices do not appear in
$\la \bar s_i\ra$  it can as well be obtained directly from  the ZRP correspondence. 
The advantage of this matrix formulation will be clear in the calculation of higher order correlation 
functions which carry these indices. 

The $2$-point correlation functions $G^{(2)}_j = \la \bar s_i  \bar s_{i+j}\ra$ are written   as   
\bea
G^{(2)}_j(z,\gamma) ={\gamma^{j+1} \over \c Z } \bra \beta   C^{j-1} 
\ket \alpha \bra \beta\frac{1}{{\cal I}-\gamma C}\ket \alpha 
= \gamma^j (1-\rho) \bra \beta   C^{j-1} \ket \alpha. 
\label{eq:G2}
\eea
All other $(n+1)$-point correlations can be expressed in terms of $G^{(1)}$ and $G^{(2)}_j$s, 
$$
G^{(n+1)}_{j_1\dots j_n}=\la \bar s_i \bar s_{i+j_1}\bar s_{i+j_1+j_2} \dots \bar s_{i+j_1+\dots j_n}\ra  
      = G^{(1)} \gamma^n \prod_{i=1}^n \bra \beta (\gamma C)^{j_i-1}\ket \alpha 
=\frac{1}{{G^{(1)}}^{n-1}}\prod_{i=1}^n  G^{(2)}_{j_i}. \n 
$$

To evaluate $\c Z(z,\gamma)$  we need to know $ \bra \beta ({\cal I}- \gamma C)^{-1}$ and 
$(I- \gamma z   D)^{-1} \ket \alpha$. The later can be   obtained from  Eq. (\ref{eq:Dm}) 
and it can be shown that, 
\bea
\bra \beta \frac{1}{I- \gamma C} = \sum_{k=1}^\infty \bra k  \frac{(\gamma z)^{k-1}}{ 1- \gamma F(\gamma z)}.
\label{eq:lZzg}
\eea
Here,  $F(w)= \sum_n  w^n f(n)$   is nothing but the 
{\it single-box} grand partition function of the corresponding ZRP with fugacity $w$. 
Using Eqs. (\ref{eq:Dm}) and  (\ref{eq:lZzg})  in (\ref{eq:Zzg})  we have
\bea
\c Z(z,\gamma)= \frac{\gamma [ F(\gamma z) + \gamma z F^\prime(\gamma z)]}{ 1- \gamma  F(\gamma z)}.
\label{eq:Zclosed}
\eea
The  advantage of  the use of VLE is clear from the above  simple and closed form expression of the partition function
that we obtain for the lattice system.  Equation (\ref{eq:lZzg}) can also be used  along with Eqs. (\ref{eq:Zclosed}) 
and  (\ref{eq:G1}) to obtain the average density in VLE,
\bea
\rho(z,\gamma) = \frac{\gamma z F^\prime( \gamma z )}{  F( \gamma z ) +\gamma z F^\prime( \gamma z )}
\eea

Higher order correlation functions can be calculated from $G^{(2)}_{m}$  with  arbitrary $m$ for 
which one requires 
\bea 
\bra \beta C^m \ket \alpha &=&\{ \gamma^m \}  \bra \beta\frac{1}{{\cal I}-\gamma C}\ket \alpha  
= \{ \gamma^m \}  \frac{F(\gamma z)}{1 - \gamma F(\gamma z)}.\n   
\eea
Here, $\{ x^n \} g(x)$ denotes the coefficient of $x^n$ in the series expansion of $g(x)$.  
To proceed further $F(w)$ needs to be specified.  In the following, we illustrate this formalism 
with some examples. 

Our first example is  the restricted exclusion process (RASEP) \cite{RASEP} defined on a  one 
dimensional periodic lattice with hardcore particles   where a  particle is  allowed 
to move to a  neighbouring vacant site, say to its right,  if it is preceded
by  at least $\mu$ particles. 
It has been shown, that RASEP and several variations of it  undergo a phase transition from 
an absorbing  state to an active phase as the density is increased beyond a critical value $\rho_c= \frac \mu{1+\mu}$.
Exact correlation functions  for RASEP are obtained by using a generalization of MPA.  
Let us reproduce these results using the procedure sketched  here, as these models can be mapped to  
a ZRP \cite{Jain} with  $u(n) = \Theta(n-\mu)$, where 
$\Theta(x)$ is the Heaviside  theta function.

The simplest case  $\mu=1$ corresponds to the  dynamics $110 \to 101$, which is 
also a special case of some other models  studied earlier \cite{earlier}  in different contexts. 
In  the following we consider  only this case; generalization to $\mu>1$  is 
straightforward. 
Here, the single box weight is  $f(n)=\Theta(n)$ and correspondingly $F(w)= {w \over 1-w}$. Thus, in this case, $\ket \alpha = (0,1,1\dots)^T.$ 
Since $\bra \beta \alpha\ra=0$, weight of any configuration with two or more consecutive $0$s 
vanishes identically, as observed  in the steady state of RASEP.

Using Eq. (\ref{eq:Zclosed}), the partition function  of RASEP is given by
\bea
\c Z(z,\gamma) = \frac{\gamma^2 z (2- \gamma z)}{(1- \gamma z- \gamma^2 z)(1- \gamma z
)}.\n 
\eea
We have checked that $\{\gamma^L\}\c Z + z^L$  is the GCE partition function  obtained 
earlier \cite{note3} using MPA. The extra factor $z^L$ is the weight of the 
configuration $\{s_i=1\}$ which does not have a ZRP equivalence.
Clearly $\c Z$  has to be  supplemented by Eq. (\ref{eq:LN}). 
For any given $z$, the average system size $\la L \ra$ can be varied by
varying $\gamma$ and the thermodynamic limit $\la L \ra\to \infty$ is reached 
as  $\gamma$ approaches 
$\gamma^*={\sqrt{z^2+4 z} -z \over 2 z}.$ 
In this limit
$
\rho(z)= {1 \over 2- z \gamma^*} =   \frac12 \left[1+\sqrt{\frac{z}{4+z}}\right], 
$ 
and the $(n+1)$-point correlation function  
\bea
  \la  s_i s_{i+1}\dots s_{i+n}\ra =  \sum_{L=n+1}^\infty \gamma^L Tr[ D^{n+1} C^{L-n-1}]  
= \frac{[z\gamma^*]^n}{2- z \gamma^*} = \rho \left[\frac{2\rho-1}{\rho}\right]^n.  \n
\eea
Both the density-fugacity relation  and the correlation functions,  calculated in VLE, are same  
as those obtained earlier in GCE \cite{RASEP}.

We close  the discussion on RASEP with the  comment that an alternative finite dimensional representation  
 $D= \pmatrix{0&1\cr 0&1}$ , $\ket \alpha = \pmatrix{0\cr 1}$  and   $\bra\beta = (1~0)$ which satisfy  
Eq. (\ref{eq:main}) is identical to the matrices  obtained in \cite{RASEP} using  MPA.  

As a second example we study another one dimensional system \cite{EKLS} where particles have extra degrees of freedom, namely spin. Each site $i$  of the periodic lattice is either vacant or occupied with 
one $+$ or $-$ kind of particle; correspondingly  $\tau_i = 0, \pm 1.$ These  particles interact  through 
an Ising-like interaction $H= -\epsilon/4 \sum_i \tau_i \tau_{i+1}$ which is reflected in the dynamics: 
\bea
+- \mathop {\longrightarrow}^{1-\Delta H} -+  ~~~;~~~
+0 \mathop {\to}^\alpha 0+  ~~~;~ ~~
0- \mathop {\to}^{\alpha} -0.   
\eea
This  model, hereafter referred to as EKLS, is an extension of an exactly solvable system studied 
by Katz-Lebowitz-Spohn \cite{KLS}. It has been shown \cite{EKLS} that  EKLS can be mapped to ZRP
approximately,  ignoring the spin indices, $i.e.$, $0$s  considered as boxes and  both $+$s and $-$s 
are considered as particles. The new site variables are now $s_i=\tau_i^2$. 
ZRP hop rate for a system having same numbers of  $+$ and $-$ particles is
 \bea
 u(n) = J_n =   \frac{v +\epsilon}{v^3} \left(  1 + \frac{b(\epsilon)}{n }\right);    
 ~b(\epsilon)=  \frac{3 v (2+\epsilon) + 6\epsilon}{4(v+\epsilon)};~
 v= \sqrt{ \frac{1+\epsilon}{1-\epsilon} }+1 ~~ \n
\eea
where $J_n$ is the current through a block  of size $n$ to the leading order in $1/n$.
The ZRP mapping is known to be exact \cite{AHR0} for the special case  $\epsilon=0$.

 \begin{figure}[h]
 \centering
\includegraphics[height=4.5 cm]{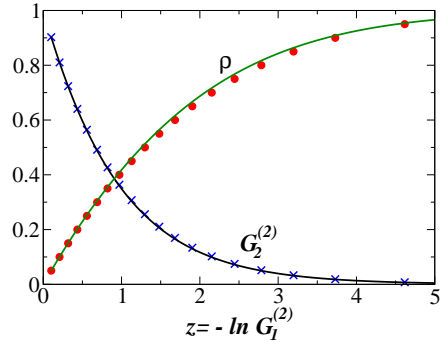}
 \caption{(Color online) $\rho$ and $G^{(2)}_2$ as a function of $z=-\ln G^{(2)}_1$.  
Results from numerical simulations (symbols) of EKLS  for $\epsilon=-8/17$  with $L=1000$ 
and $\alpha=1$ are compared with theoretical calculations (solid line).}
 \label{fig:1}
\end{figure}

As $\epsilon$ is increased beyond a critical  value $4/5$, where $b=2$, a phase separation transition  occurs in  EKLS model for
large densities. Extensive numerical studies have provided evidence \cite{EKLS} of 
such a transition. However, to the best of our knowledge, the  spatial correlations of 
these models have  not been studied yet.  Let us use  the formalism discussed here to 
calculate these correlations and compare  them with  those
obtained from numerical simulations.  Although the spatial correlations can be found 
for  any arbitrary $b(\epsilon)$, we choose to work with $b=1$ corresponding to  $\epsilon=-8/17$ 
(antiferromagnetic interaction) where it is possible to find simple closed form expressions. 
Then ignoring the  prefactor, which only 
redefines the fugacity, we have   $u(n) = (n+1)/n$. Thus $f(n) = 1/(n+1)$  and 
correspondingly $F(w) =-ln(1-w)/w.$ 
The partition function  takes a simple form 
$$\c Z = \frac{\gamma z}{(1 - \gamma z) (z + \ln(1 - \gamma z))}.$$
In the thermodynamic limit $\la L \ra\to \infty$, which  corresponds to $\gamma^*= z^{-1}(1-e^{-z})$,  
the density  is $ \rho= 1-z/(e^z-1).\label{eq:kls_rho_z} $  The two-point 
correlation functions (\ref{eq:G2}) are then, 
 $$
G^{(2)}_j(z)= {\gamma^*}^j (1-\rho) \sum_{i=1}^j \frac{z^j}{ i (j-i)!} \int_0^\infty d\lambda  e^{-\lambda z} (\lambda)_{j-i},
$$ 
where  $(\lambda)_{l}= \lambda (\lambda +1)\dots (\lambda +l-1)$ is the Pochhammer symbol. 
Explicitly for $j=1,2$
 $$
G^{(2)}_1 =  e^{-z}, ~~ G^{(2)}_2 = \frac{ (1-e^{-z})(2+z)}{2 z e^z}. \n 
$$

Usually the correlation functions in GCE are  compared with those in the
canonical ensemble (having fixed density $\rho=N/L$) by expressing $z$ in 
terms of $\rho$.  However, here   it is simpler to use  $z= -\ln G^{(2)}_1$  and 
express  $\rho$ and other correlation functions  in terms of $G^{(2)}_1.$
In Fig. 1 these results are compared with 
the $2$-point correlation functions  obtained from the numerical simulations of EKLS with 
different densities $\rho$. An excellent match indicates 
that the mapping of EKLS to ZRP is a good approximation.

In summary, we have studied an exclusion process  on a one dimensional ring
where a particle moves to  its rightward vacant neighbour with a rate that depends on the 
size of the block  to which the particle belongs. The model has a natural correspondence to  
zero range process but positional ordering of the particles is lost in this mapping.
Thus the spatial correlations of the exclusion process are  
unreasonably difficult to calculate  directly from the
steady state weights of ZRP. A  method is introduced to obtain the same,  by rewriting the 
 steady state weights of the exclusion process in matrix product form.  The  required matrices are constructed  explicitly using only the single box weight of the corresponding ZRP. The method 
is illustrated using the  example of  RASEP.
Since, several one dimensional exclusion processes are known 
to have ZRP correspondence,  this formulation can be used to study  unexplored spatial correlations 
in such systems. As immediate applications, spatial correlations   
in EKLS model (discussed here), Tonk's gas and  RASEP with parallel dynamics 
(to be discussed elsewhere) have been studied. We believe that this method is a useful tool  
for the study of  correlations in one dimensional non-equilibrium steady states. 

We are grateful to Deepak Dhar for his  useful comments on the manuscript.
U.B. would like to acknowledge thankfully the financial support of the
Council of Scientific and Industrial Research, India (Grant
No. SPM-07/489(0034)/2007).


\begin{thebibliography}{99}
\bibitem{note3} Note that in  \cite{RASEP} the fugacity was taken as $z^2$.
\end{thebibliography}
\end{document}